\documentclass[sigconf]{acmart}

\AtBeginDocument{%
  }

\copyrightyear{2025}

\acmConference[Preprint]{Preprint}{2025}{arXiv}

\usepackage{csquotes}
\usepackage{xcolor}
\usepackage{enumitem}
\usepackage{dblfloatfix}
\usepackage{graphicx}

\begin{document}

\title{Aging Up AAC: An Introspection on Augmentative and Alternative Communication Applications for Autistic Adults}


\author{Lara J. Martin}
\authornote{Work done while at the University of Pennsylvania.}
\email{laramar@umbc.edu}
\affiliation{%
  \institution{University of Maryland, Baltimore County}
  \city{Baltimore}
  \state{MD}
  \country{USA}
  \postcode{21250}
}
\author{Malathy Nagalakshmi}
\email{malathyn@seas.upenn.edu}
\affiliation{%
  \institution{University of Pennsylvania}
  \city{Philadelphia}
  \state{PA}
  \country{USA}
  \postcode{19104}
}

\renewcommand{\shortauthors}{Martin \& Nagalakshmi}
\renewcommand{\shorttitle}{Aging Up AAC}

\begin{abstract}
High-tech Augmentative and Alternative Communication (AAC) has been rapidly advancing in recent years due to the increased use of large language models (LLMs) like ChatGPT, but many of these techniques are integrated without the inclusion of the users' perspectives. Autistic adults have been particularly neglected in the design of AAC tools. We conducted in-depth interviews with 12 autistic adults to find the pain points of current AAC and determine what technological advances they might find helpful. We found 8 different categories of themes from our interviews: input flexibility, output flexibility, selecting or adapting AAC, contexts for AAC use, benefits, access as an adult, stumbling blocks for continued use, and control of communication. In this paper, we go through these categories in depth -- comparing each to prior work -- and then highlight novel findings to suggest possible research directions.
\end{abstract}


\begin{teaserfigure}
\centering
  \includegraphics[alt={Figure 1 is a collage of various AAC applications used by participants. In the collage there is:   1. A picture of a hand holding a phone that shows the app My Voice.   2. A screenshot of someone's homeboard in CoughDrop.   3. A picture of someone typing on a keyboard in front of a monitor with a tablet in their lap.   4. A screenshot of someone typing a sentence in Proloquo.   5. A picture of a tablet showing TD Snap.},width=.8\textwidth]{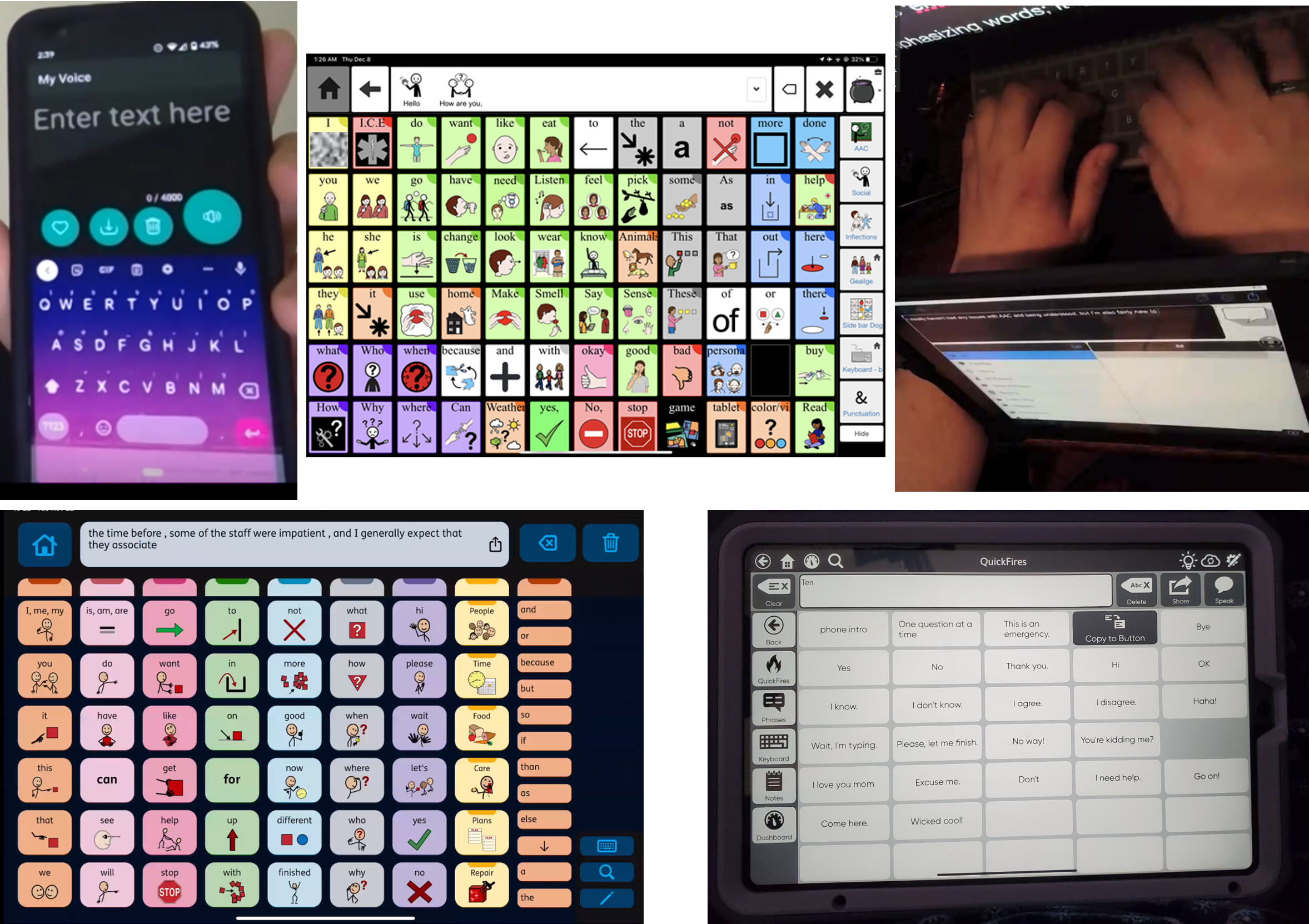}
  \caption{A compilation of our participants' various AAC devices and interfaces. The applications (from top to bottom, left to right) are My Voice, CoughDrop, ClaroCom (with keyboard setup), Proloquo, and TD Snap.}
  \label{fig:teaser}
\end{teaserfigure}

\received{16 April 2025}

\maketitle

\section{Introduction}

Thanks to the neurodiversity movement online \cite{Kapp_2020,Botha_etal_2024}, many people who were previously overlooked for diagnoses as children are now realizing that they are autistic as adults.
Since these autistic individuals come into their diagnoses well into adulthood, they are learning about accommodation much later in life \cite{Donaldson2021}.
This means that tools like Augmentative and Alternative Communication (or AAC), which replace or supplement a person's oral communication, are discovered by people who 1) already know how to read and write, 2) have been speaking orally (to their detriment) for years \cite{Zisk2019}, and 3) may not have access to professionals such as speech-language pathologists. 
Furthermore, the tools themselves are usually developed with children in mind, with speech-language pathologists teaching children how to use AAC to speak their first language.

Although there has been recent recognition within AAC research showing that AAC use can build agency \cite{Valencia2020} and garner empathy from people in the AAC user's life \cite{vanGrunsven2022}, 
autistic adults are largely ignored within AAC research \cite{Ganz_Earles_2012,Trembath2014,Lorah2022}, especially when designing new AAC tools \cite{Bastable2021}.
Because of this, autistic adults have learned to navigate a world of AAC tools that are not made for them.


Technologically-speaking,
research on high-tech AAC tools\footnote{AAC tools can range across the spectrum of technology levels---from writing, spelling out words on a letter-board, or using cards with pictures on them; to tablet/phone-based applications that cost several hundreds of dollars.} has seen a recent uptick in interest from the Natural Language Processing (NLP) community due to the introduction of large language models (LLMs) like ChatGPT to the general public \cite{cai2022context,Kumar2022,valencia_less_2023,yusufali2023bridging}. However, we should tread carefully as NLP and other machine learning techniques can cause harm if not implemented correctly \cite{Huang_2023}. 
There are crucial privacy issues such as
training on users' data without their knowledge \cite{Crawford_2024} or sending/saving data to ``the cloud'' with questionable levels of security \cite{Hamidi_2018}.
Furthermore, implementing NLP techniques blindly without the inclusion of the community of users can backfire. 
Thankfully, there has been a push to include AAC users as contributors in AAC development \cite{kudryashov2021participatory,Bastable2021,Beneteau2020,Taylor2020}. Our work continues down this path by beginning conversations with AAC users to \textbf{bridge the gap between what can be done technologically and what users actually want}, with a focus on high-tech, computer-based AAC applications.

In this paper, we characterize the needs of autistic adults in order to improve AAC applications and investigate deeper issues as to why some adults do not use AAC applications.
In the vein of looking to improve AAC tools for a sub-community of users who are often unheard---autistic adults, we were interested in the following research questions:
\begin{itemize}
    \item \textbf{RQ1:} What are the benefits and challenges that autistic AAC users face when using existing AAC applications?
    \item \textbf{RQ2:} What types of features do autistic adults require or prefer to see from AAC applications?
    \item \textbf{RQ3:} How does a person's identity as an autistic person factor into their AAC use?
    \item \textbf{RQ4:} How do autistic adults feel about various technical considerations for future AAC applications (i.e., the integration of NLP techniques)?
\end{itemize}

To learn more about this intersection of identities, we conducted 12 semi-structured in-depth interviews with autistic adults who were recruited online. 
We outline themes that we have identified throughout the interviews and compare our findings to prior work \footnote{Many social issues regarding AAC use arose in addition to technological issues. We will not be discussing the social issues since they are outside the scope of this paper. If this is relevant to your work and you would like to know what we found, please contact the first author.}.
Finally, we highlight potential new research directions based on our interviews.

To match the language autistic adults use \cite{Zisk2022}, we will be using the terms "speaking orally", "using mouth-words", or "speaking" (vs. "nonspeaking") interchangeably throughout the paper to refer to using one's vocal tract to produce language.


\section{Previously-Identified Issues with AAC}
\label{sec:related}

\subsection{Autistic Adult AAC Users}

Some work with intellectual and developmental disabilities broadly, including people with autism spectrum disorder, has looked at how AAC can improve users' skills such as single word sight reading \cite{Holyfield2020} or achieving their own personalized communication goals \cite{Hughes2022}.
These types of work track how well participants \textit{use} AAC rather than how they feel about the technology.
To our knowledge, there have only been two papers written so far looking specifically at autistic adults' general AAC use \cite{Zisk2019,Donaldson2021}---both papers focused on speaking adults. Our study mirrors these works since we interviewed autistic adults who were not ``completely nonspeaking''. (We speculate that it is difficult for someone to consider themselves fully nonspeaking if they have forced speech for most of their life and only encountered AAC as an adult, as many of our participants have.)

\citet{Zisk2019} reviewed academic and non-academic sources to find recurring themes of how speaking autistic adult AAC users communicate when they can occasionally use speech, what types of tools they use to communicate (both specialized for AAC and not), and any barriers to using AAC. 
They mention that autistic adults are unaware of AAC options or ways of getting support, and that lacking an autism diagnosis can impede adults from getting the aid they need. \citet{Zisk2019} also identify the cost of applications or preconceived notions of AAC as issues for autistic adults to acquire AAC.
\citet{Donaldson2021} performed an open-ended written survey with six participants, asking questions specific to AAC history, access, and attitudes toward AAC use. They also asked participants for any recommendations for people who might use AAC or interact with people who use AAC. While our work has rediscovered some of the same social themes as \citet{Donaldson2021} (e.g., choice of AAC, autonomy), we also dive into technical and privacy themes. \citet{Donaldson2021} also focus more on discovering AAC as an adult, acceptance of AAC use, and forcing speech.
These papers will be referenced throughout this work.

\subsection{Other Populations of AAC Users}
There are some fairly universal issues with AAC use among adult users.
Some recurring themes across the literature include issues with the speed of communication and conversation pacing \cite{higginbotham2016time,Kane_2017,Waller2019,Shin_2020}, access to AAC \cite{Waller2019,Dai2022,Amery2022,Pope2022,Donaldson2023}, being sufficiently trained in using the AAC \cite{Waller2019,Gibson2020}, low-quality speech synthesis \cite{Beukelman_2007,Kane_2017,Gibson2020}, and difficulties finding words \cite{Shin_2020,Obiorah_2021}. Past participants mentioned a tendency for AAC devices to be too focused on needs rather than other topics of conversation \cite{Kane_2017,Waller2019,Obiorah_2021,Dai2022} or lack personality/humor \cite{Kane_2017,Dai2022,Weinberg_2024}.
There are also impacts on the agency (both positive and negative) of the AAC user \cite{Ibrahim_2018,Valencia2020,Hajjar2022,Donaldson2023}.
Compared to these past works that focus primarily on the pain points of existing technologies, we ask users to reflect on issues of existing AAC technologies while also considering how future AAC might be made. In particular, we orient participants to think about empowerment, NLP/AI use, and privacy.

There have also been studies including specific populations of non-autistic (i.e., allistic) adults who use AAC.
In addition to issues mentioned in the previous paragraph, \citet{Gibson2020} found adults with mild intellectual disabilities want customizable interfaces, tutorials on how to use the app, fewer options presented at a time, images to explain button functionality, and audio playback of text.
\cite{Obiorah_2021} saw that adults with aphasia would like to integrate other digital devices into their AAC use.
\citet{Kane_2017} worked with adults with ALS, who experienced difficulties with volume and switching roles in a conversation.

Very little work \textit{directly} involves children with disabilities in the design of assistive technologies \cite{Benton_2015,Spiel_2019}. Instead, much of the work has a clinical focus---looking at how effective the child's communication is \cite{Logan_2017}, or asking parents/guardians, teachers, etc. of the children to describe the child's AAC use \cite{Beneteau2020}.
That said, \citet{Shin_2020} have noted that autistic children have trouble generating full sentences with their AAC and their parents have difficulty modeling sentence creation with symbols. \citet{Walker_2022} saw that autistic children more often use AAC with adults than with other children.
Most autistic children have a preference for high-tech AAC (vs low-tech AAC or signing) \cite{Lorah2022}.
Throughout this paper, we also mention previous work relevant to the corresponding themes.


\section{Methods}

We setup an in-depth interview to ask autistic adults how AAC applications can be improved. This resulted in six sets of questions. The questions were then verified by two autistic researchers who specialize in qualitative work who were not affiliated with the study. The full set of questions can be found in Appendix \ref{sec:appendix-speaking}, but the sets are summarized as follows:
\begin{enumerate}
\setlength\itemsep{0em}
\item Questions asking the participant to recall a recent conversation where they used AAC and describe details about how that conversation went (RQ1).
\item Questions asking the user to think of particular scenarios when they used AAC (RQ1).
\item Questions for describing a typical day of using AAC and what types of tools they used (RQ1,RQ2).
\item Questions asking about when they have felt empowered or disempowered because of their AAC (RQ3).
\item Questions about how they select what AAC to try (RQ2).
\item Questions about privacy (ending with a privacy table) (RQ4).
\end{enumerate}
The privacy table was a list of potential NLP features that were presented to the participant after thinking about the privacy of using AAC on their own. For each of the seven features we came up with, the participant had to rate whether they would allow it in an AAC application they would use. More information about the table and its results is presented in Section \ref{sec:control}. The research design, as well as the consent form, were approved by the IRB board of the University of Pennsylvania.

We recruited participants via the Facebook group ``Ask Me, I’m an AAC user!'' and X (formerly Twitter).
The participants were asked to fill out an initial intake questionnaire (Appendix Section \ref{sec:appendix-intake}), and if they qualified, they were emailed the consent form and a video of the first author reading the consent form aloud\footnote{This practice of recording the consent form was taken with permission from \href{mailto:CXB1058@student.bham.ac.uk}{Charlotte Brooks}, which was well-received by autistic adult participants.}.
If they consented, they were asked to schedule an interview and were given a PDF containing all the interview questions \& a PDF with the privacy table---the latter of which they were told not to open until the appropriate time. 
Participants were allowed to script their answers to the questions ahead of time, and were especially encouraged to do so if they had planned to use AAC during the interview.

We kept the recruiting open until we had 12 participants \footnote{It has been shown \cite{Ando2014} that themes begin to stabilize after around 12 participants.}.
Seventy-eight people signed up for the study and filled out the initial intake questionnaire, twenty-five people were invited to participate in the interview, and twelve people were interviewed.
For confidentiality, we have revised the anecdotes to eliminate any information that could potentially identify the participants.

All participants were native English speakers, were living in the United States, were between the ages of 18-44 years, and used some form of high-tech AAC at least occasionally. Only two participants (2) said they used AAC full-time; the rest said that they were part-time users. A list of each participant's age range, self-identified speaking label, current AAC, and amount of time using AAC (both in total and for the current AAC) can be found in Table \ref{table:participants}.

Most of the interviews were performed via Zoom with video and audio. The interviews lasted about an hour, with the average being 1 hour and 18 minutes (standard deviation of 31.2 minutes, minimum 50 minutes, maximum 2 hours \& 40 minutes). Interviewees were encouraged to use their AAC during the interview. One interview was conducted entirely asynchronously via email \& text document. Another interview was conducted half on Zoom (cut off at 1 hour) and half asynchronously.
For consistency, all interviews were run by the first author. 

\subsubsection{Analysis}
We analyzed the interviews through thematic analysis \cite{ThematicAnalysis}. We used Zoom's built-in automated transcription tool to get an initial draft of the transcript. The transcripts were then corrected and verified by the 
authors. If an important section of the interview could not be confidently transcribed, the authors asked the interviewee for clarification via email. The authors independently formulated an initial set of themes by analyzing the transcripts in an asynchronous manner. Subsequently, they engaged in discussions to merge their respective themes, followed by a process of refining these combined themes and evaluating their alignment with the transcript content. Through iterative revisions and affinity diagramming, the authors reached a consensus on the final set of categories and themes presented in this paper.

\section{Categories of High-Tech AAC Applications}

We observed a divide between two categories of AAC applications based on input type: those that use typing-based input and those that use symbol-based input. Typing-based apps enable users to input any word they want on the fly, while symbol-based apps require pre-programming of what words will be displayed on buttons and what the layout of the buttons will be. The divide seems to revolve around two commonly-used applications (\textit{Proloquo2Go} and \textit{Proloquo4Text}), both developed by the company AssistiveWare\footnote{\href{https://www.assistiveware.com/}{https://www.assistiveware.com/}}. 
During the course of this research, AssistiveWare had come out with a new application in the Proloquo suite simply called \textit{Proloquo}. \textit{Proloquo} seems to solve some of the problems of this divide by beginning to merge the functionality of \textit{Proloquo2Go} and \textit{Proloquo4Text}. \textit{Proloquo} is now a button-based layout similar to \textit{Proloquo2Go} but without symbols (only words) and with more predictive text options. \textit{Proloquo4Text} consists primarily of on-screen typing with the option to save phrases.

Other commonly-used applications among our participants include \textit{CoughDrop}\footnote{\href{https://www.coughdrop.com/}{https://www.coughdrop.com/}} by CoughDrop, Inc. and \textit{TD Snap}\footnote{\href{https://us.tobiidynavox.com/pages/td-snap}{https://us.tobiidynavox.com/pages/td-snap}} by Tobii Dynavox---which are both symbol-based apps. 
We have seen that those who prefer typing or swipe on phones \& tablets usually find free applications that will support them in the ways they need.
Please see Table \ref{table:participants} for the full list of applications and tools that our participants use to communicate.

\begin{center}
\begin{table*}
\begin{tabular}{ |p{2.5em}|p{15em}|p{7em}|p{15em}|p{4em}| } 
\hline

\textbf{Age Range} & \textbf{Self-Assigned Label} & \textbf{Years \newline (Total)} & \textbf{Current AAC} & \textbf{Years \newline (Current)}  \\\hline

18-24 & Mostly non-speaking &  5-9 &\textcolor{red}{Proloquo2Go}, \newline \textcolor{red}{Proloquo4Text}, \newline \textcolor{red}{TD Snap} (for Swedish)& 2-4 \\\hline

 25-34 & Unreliable speech & since childhood &\underline{writing},\newline \underline{ASL (working proficiency)},\newline {UbiDuo} (device),\newline \textcolor{red}{BuzzCards},\newline Relay & 5-9 \\\hline

 25-34 & Semi-speaking &  5-9 & \textcolor{red}{Proloquo4Text}, \newline \textcolor{red}{Proloquo2Go},\newline {\color{red}Tell Me}& $\le$ 1  \\\hline

 25-34 &Mostly non-speaking & $\le$ 1&\underline{\textcolor{red}{Proloquo}}, \newline \textcolor{red}{Proloquo4Text},\newline Relay,\newline \textcolor{red}{CoughDrop}&$\le$ 1\\\hline

 25-34 & Selectively mute/situationally mute &  10+ &\underline{texting/messaging}, \newline\textcolor{red}{Speech Assistant},\newline \textcolor{red}{Proloquo2Go}, \newline writing& 2-4 \\\hline

 25-34 & Semi-speaking &  10+ &\textcolor{red}{CoughDrop},\newline \textcolor{red}{Predictable},\newline \textcolor{red}{Speech Assistant},\newline many low-tech options& 2-4\\\hline

  25-34 & Semi-speaking & $\le$ 1 &\textcolor{red}{My Voice},\newline email,\newline Relay& $\le$ 1 \\\hline

35-44 & Mostly non speaking [with] unreliable speech and word retrieval issues & 10+ &\underline{\textcolor{red}{Proloquo2Go}}, \newline  \underline{\textcolor{red}{TD Snap}}, \newline \textcolor{red}{CoughDrop}& 2-4\\\hline

35-44 & Sometimes-user &  5-9 & \underline{\textcolor{red}{Proloquo4Text}},\newline \textcolor{red}{Proloquo2Go},\newline Relay& 2-4  \\\hline

35-44 & Mostly non-speaking &  $\le$ 1 &\textcolor{red}{ClaroCom},\newline {\color{red}Proloquo}& $\le$ 1 \\\hline

 35-44 & Mostly non-speaking &  5-9 & \underline{\textcolor{red}{Proloquo2Go}},\newline {\color{red}Proloquo4Text}& 5-9 \\\hline

 35-44 & Unreliable speech &  $\le$ 1 &\textcolor{red}{TD Snap},\newline \textcolor{red}{Speech Assistant}& $\le$ 1 \\\hline
\end{tabular}
\caption{List of participants with their age range, self-assigned speaking label, years spent using AAC, their current AAC applications (the \underline{underlined} app/technique being what they mostly use), and the number of years spent with their current application. Words in \textcolor{red}{red} are phone/tablet applications. ASL stands for American Sign Language. Relay is a free federal service for communicating with text over the phone.}
\Description{Table 1 is a list of participants and their varying AAC use. Most AAC applications listed are in red (denoting applications), with the exception of writing, ASL, UbiDuo, Relay, texting/messaging, email, and many low-tech options. Some participants have AACs that are underlined (denoting what they use the most), but others don't have a most-commonly-used AAC.
}
\label{table:participants}
\end{table*}
\end{center}

\section{Results}

This section delves into the categories identified during the qualitative analysis of the transcripts. We use a numerical value enclosed in parentheses to indicate the number of participants supporting that particular theme (e.g., "(3)" signifies "three interviewees"). We will not be using any individual IDs due to the close-knit community of AAC users. Because of this, we will refer to everyone by "they/them", although we collected their pronouns during the interview. Direct quotes will be surrounded by quotation marks and italicized. The complete list of all 12 participants can be found in Table \ref{table:participants}.

We have identified 9 main categories of themes.
Because our participants had a lot to say about what they look for in an AAC application (RQ2), we split this into three categories: 
\textit{Input Flexibility} (\S\ref{sec:input}),
\textit{Output Flexibility} (\S\ref{sec:output}),  and 
\textit{Selecting or Adapting AAC} (\S\ref{sec:selecting-adapting}).
\textit{Benefits} (RQ1) can be found in Section 
\ref{sec:benefits}. Challenges (RQ1) can be found in \textit{Stumbling blocks for continued use} (\S\ref{sec:stumbling}).
For RQ4, we have the category: \textit{Control of Communication} (\S\ref{sec:control}).
Themes surrounding our participants' identities (RQ3) can be found across categories.
Additionally, two categories emerged from the interviews:
\textit{Contexts for AAC Use} (\S\ref{sec:swap}) and
\textit{Access as an Adult} (\S\ref{sec:access}).

\subsection{Input Flexibility}
\label{sec:input}

All 12 participants had preferences of what input type they preferred or would like to see in future AAC applications. In this section, we discuss the use of keyboards (\S \ref{sec:typing}), the differences felt between symbol-based and text-based AAC (\S \ref{sec:symbol-text}), and aspects that make input easier such as vocabulary (\S \ref{sec:vocab}) \& layout flexibility (\S \ref{sec:layout}), predictive text/pre-programmed phrases (\S \ref{sec:predictive}), and button selection interfaces (\S \ref{sec:indirect}).

\subsubsection{Typing} 
\label{sec:typing}
\begin{figure}[h]
\includegraphics[alt={Figure 2 is a picture of an iPad with the program Proloquo2Go. The program shows a grid of buttons with varying symbols denoting words. On the bottom of the grid there is a QWERTY keyboard laid out in yellow buttons (to visually pop compared to the usually-white button backgrounds).},width=.45\textwidth]{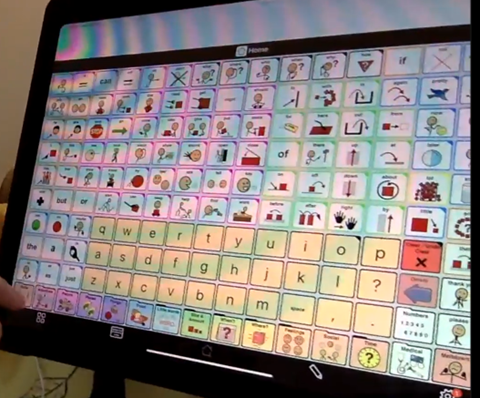}
\centering
\caption{One participant showing their on-screen keyboard embedded on the homepage in \textit{Proloquo2Go}.}
\label{fig:embedded-keyboard}
\end{figure}

Since our participants were recruited because they were autistic, they all had varying degrees of motor control, which often changes depending on their physical, emotional, or mental state. 
That said, all participants (12) wanted some type of typing input option to use at least occasionally. This could range from full-time use to simply augmenting their vocabulary for when they don't have an appropriate button.
Although a few (3) prefer a physical keyboard (``\textit{I can type on a phone, but I type 100 words a minute on a regular keyboard so that's always gonna be my preference.}''), most prefer or at least default to using an on-screen keyboard (8). One person (1) would take any kind of keyboard access ``\textit{even if it's an on-screen keyboard}'' but finds a physical keyboard better for their wrists and hands.
Within those who prefer on-screen keyboards there was some variation in preferences of how that keyboard is implemented.  Some (3) prefer a keyboard embedded within the application itself. One participant (1) said this is preferred because an OS-native on-screen keyboard would mean ``\textit{needing to swap motor plans back and forth a lot}'' (1)\footnote{Motor planning is the process of learning how to perform an action or series of actions---or as one participant put it: ``\textit{Your body knowing how to body.}''}. A couple participants (2) had gotten around this by embedding a keyboard in the homepage of their AAC as regular buttons (for example, Figure \ref{fig:embedded-keyboard}).

\subsubsection{Symbol- vs text-based AAC}
\label{sec:symbol-text}
On the other hand, some participants have a preference for symbol-based AAC.
One person (1) said that symbol-based apps work better for their brain but are much slower.
Another person (1) said that they prefer having some images, although it is not a requirement.
A few participants (4) explicitly mentioned the difference they feel when using word- vs symbol-based AAC. They find typing to be faster (2) but more taxing (2) --- creating ``\textit{more mental effort}'' or being ``\textit{less calm}''. Meanwhile, regarding symbol-based AAC, one participant said ``\textit{there's less pressure for me to find the correct words,}'' while another said 
``\textit{it forces me to slow down and think about what I'm trying to say.}'' 
Others will actively avoid symbol-based AAC (2), or allow buttons as long as they don't have symbols (1).
\citet{Holyfield_2021_comparative} has shown that even pre-literate autistic children take about the same amount of time to learn symbol-based or text-based AAC, so perhaps the choice is personal preference.

Some participants (7) suggested supplementing or mixing symbol usage with typing, which can include things like 
 using symbol-based AAC with Zoom's chat (1) or completely mixing the two types of AAC (``\textit{Future AAC that has both symbols and typing and a vocabulary designed for autistic adults would be very empowering.}'').
Having applications focus solely on symbol-based \textit{or} text-based input creates unnecessary friction:
\begin{displayquote}
``\textit{[I dislike] symbols and typing being so completely separate in different apps.
Having multiple apps is good and choice is good but [...] I don't want to have to keep switching,
and the typing in the symbol app just isn't sufficient for what I need.}''
\end{displayquote}

\subsubsection{Vocabulary} 
\label{sec:vocab}
One downside to using symbol-based AAC is that it could be limiting in the amount of vocabulary available.
As adults who have been using mouthwords for years, our participants have very large vocabularies. A couple of participants (2) have had experiences with older AAC with severely limited vocabulary. Seven of our participants (7) look for applications with high flexibility in terms of vocabulary, which may include an adult-level vocabulary (2), vocabulary for specific groups of people (2), 
or an overall comprehensive lexicon (7).
The integration of adult or job-specific vocabulary has also been suggested by \citet{McNaughton_2007}.

\subsubsection{Layout} 
\label{sec:layout}
There were a lot of different preferences for how users want their buttons to be laid out. Some of these our participants wish existed in their preferred app or wish existed in general, others already exist and could be the reason why they chose the app that they use. Our participants (7) had different opinions regarding aspects of the app's grid size (4), word organization (5), ease of use (4), button size (2), quality of the core board (2), and ability to use the same layout across applications/devices (2).

\subsubsection{Predictive \& pre-programmed text} 
\label{sec:predictive}
Participants also enjoy input that aids them with writing what they want to say, whether it be auto-complete/word suggestion (4), or helping them find the button they are looking for via a search bar (2).
However, our participants echoed some of the issues of existing AAC prediction, such as the system unhelpfully storing misspellings (1) \cite{McKillop2018} or suggesting words that were inappropriate for the social context (1) \cite{McKillop2018,valencia_less_2023}.
Thankfully, next-word prediction for AAC is frequently researched 
\cite{Schadle2004,Trnka_2007,wiegand2012non,freedman2013hierarchical,adhikary2019investigating} -- as well as contextual or discourse-based prediction \cite{freedman2013hierarchical,mitchell2012discourse,perkoff2021dialogue}, 
personalized language \cite{freedman2013hierarchical,DiMascio2019,Li2022,valencia_less_2023}, or generating sentences from a set of keywords \cite{cai2022context,Kumar2022}.
 More work is needed for determining how predictive text is displayed \cite{Yusufali_2023} and how to guarantee that the predictive text does not undermine autonomy \cite{valencia_less_2023}.

Many participants (10) also used features to pre-program whole phrases and commonly-used messages ahead of time.

\subsubsection{Indirect selection} 
\label{sec:indirect}
The ``default'' input method for symbol-based AAC appears to be \textbf{direct select}. That is, tapping directly on the button of the word the user wishes to say.
However, even in a group such autistic adults who are not necessarily using AAC due to a physical disability, some users still require or prefer an alternative mode of input. Sometimes participants (3) will switch to an \textbf{indirect select} input type (e.g., eye gaze, automatic scanning, or step scanning) because a dynamic disability such as seizures (1) or hemiparesis (1) prevents them from easily using direct select.

\subsection{Output Flexibility}
\label{sec:output}

The two main themes that popped up in terms of desired output options were having the ability to show text to conversation partners and to have customizable text-to-speech (TTS)---also called speech synthesis---that fits the user's identity.
TTS enables the user to share their thoughts with an entire group instead of being confined to showing a screen or piece of paper. 
We will go into these two themes (\textit{Showing text} and \textit{Voice identity}) in more detail in this section. (We will talk about issues with showing text and speech synthesis in Sections \ref{sec:voice-quality} \& \ref{sec:text-limit}, respectively.)

\subsubsection{Showing text} 
Many participants (8) use the text display feature of their AAC app instead of, or in conjunction with, the TTS. This can stem from conversation participants unable to understand the synthesized voice and/or because the voice couldn't be loud enough (6). Others prefer to communicate by showing text most or all of the time (3) or switch between speech and text (1).

\subsubsection{Voice identity} 
\label{sec:voice-identity}
A person's identity is strongly tied to their voice \cite{Nathanson_2017}, and having a unique, personalized AAC voice can be empowering \cite{Engelke_2013,Mills_2014,McGettigan_2015,Patel_2016,Zeffate_2023,Preece_2024}. 
Our participants expressed an interest in having access to options for different types of voices (2) or access to specific voices from voice packs that they have used in the past (2).

On top of this, many of our participants were transgender or gender nonconforming and wanted voices to reflect this.
Although one participant was happy some voice packs were beginning to include a variety of voices, a common complaint that we saw was that there are not enough nonbinary or middle-pitch voice options (4), and this becomes even more limiting when the app  
prohibits even basic customizations such as adjusting the pitch (3). The only person who uses he/him pronouns out of our participants dislikes the men's voices and prefers to pitch down child/teen TTS.
Even if we were to sample their speaking voice to create a TTS voice, it might not match what they actually want to sound like, making voices based on the user a less desirable option:
   ``\textit{
   I am trans, but I deal with age dysphoria as well, which is part of why I like [this voice]. Having a voice that sounds right is, therefore really, really important.}''

\subsection{Selecting or Adapting AAC}
\label{sec:selecting-adapting}

Beyond input/output, participants have other criteria for selecting the AAC app that is right for them. In this category, we will go over themes around applications that fit the needs and preferences of the user.
Our participants described wanting to have a single or a few AAC application(s) that were flexible in design and would have the option to be used in multiple ways to suit their needs.
This category describes themes where participants expressed the need for robust, adaptable features for their AAC so that they weren't constrained to using an app that didn't fit what they needed. These include: \textit{Aging up AAC} (\S \ref{sec:aging}), \textit{Consolidation of features} (\S \ref{sec:consolidation}), and \textit{General customization} (\S \ref{sec:customization}).

\subsubsection{Aging up AAC} 
\label{sec:aging}
\begin{displayquote}
     ``\textit{Many AAC apps feel like they're made for kids or students, and it feels infantilizing.}''
\end{displayquote}
Participants (5) were of the opinion that AAC needs to be tailored for autistic adults, especially since they are often targeted toward children.
Four participants (4) had suggestions for how AAC could be modified for adults: providing the user access to more adult vocabulary (2; also mentioned in \S\ref{sec:vocab}), which \citet{McNaughton_2007} have also found; switching from symbol-based to text-based AAC as someone grows up, or combining the two modalities (1); and
no recording of word frequency usage\footnote{Word usage is usually sent to a speech-language pathologist when learning how to use AAC to inform them ``\textit{what words I need to practice or whatever}''.}, especially since this can be seen as another way of logging data (1).
Further work should be done to study how the autistic adult community believes AAC can ``age up''.


\subsubsection{Consolidation of features}
\label{sec:consolidation}
Similar to other AAC users \cite{Obiorah_2021,Dai2022}, our participants wish AAC could fit into their existing tech ecosystem.
Three participants (3) said that consolidating features from various apps into a single platform would make things more convenient, eliminating the need to constantly switch between different applications.
There are also existing phone features, such as texting or voice calling, that participants wished could be better integrated into their AAC applications to make users' lives easier (3). 
Other participants had a dedicated device for AAC (5) or expressed interest in getting one (1).

\subsubsection{General customization}
\label{sec:customization}
All of our participants (12) expressed various levels of interest in overall customization and ease of use. This makes sense considering autistic people have different types of needs.
One of the most commonly-expressed preferences was that participants wanted their ideal AAC application to not feel overwhelming (7). Depending on the person, this could mean an interface that is not visually overwhelming (2), such as a clear layout and organization with fewer possibilities (4), or no bright colors (3)---especially when they are already feeling overwhelmed.

\subsection{Contexts for AAC Use}
\label{sec:swap}

There are a few theories about why autistic people may have trouble speaking orally. Some suggest that it could be anxiety-based \cite{Muris_Ollendick_2021,Ludlow_Osborne_Keville_2023}---although it is hard to say whether anxiety is creating the difficulties or that having difficulties speaking orally gives people anxiety.
\citet{Melfsen_Romanos_Jans_Walitza_2021}'s ``unsafe world model'' of selective mutism extends beyond anxiety and describes the lack of oral speech as a ``shut down'' of the nervous system from a perceived lack of safety.  In this model, over-stimulation (among other things) can cause the perception of being unsafe.
Related to this, two recurrent themes that came up in our study were the aspects of \textit{Trust} of other people and the use of AAC to help them with \textit{Accommodating to current needs}, which we will discuss in this section.

\subsubsection{Trust} 
Because autism is an invisible disability and some people have the ability to pass as neurotypical, using AAC can ``out them'' as autistic. Nearly all of our participants (11) spoke about how using AAC depended on whether the environment is safe, what the comfort level is with the person they are speaking to, and social safety.
\begin{displayquote}
    ``\textit{I don't use AAC [...] when there are people who have dramatic power over me who can drastically control my life, because my trauma tells me it is actively not safe to in those situations.}''
\end{displayquote}
As soon as that trust is broken, users would rather force speech than use AAC in front of them. 
\begin{displayquote}
``\textit{My friend said it felt a little incongruous to hear a child's voice from [it]. So I only used AAC with them twice.}''
\end{displayquote}
Some participants spoke about being more likely to use oral speech (3) or speak orally alongside AAC (3) with people they trust and in private situations. They are comfortable transitioning between speech and AAC in the presence of those who create a comfortable environment for communication.

Other participants do not trust strangers to accept their AAC well due to being treated poorly in the past (3), 
and can feel pressured to match others' communication expectations (2). \citet{Donaldson2021} also found trust to be a major theme influencing autistic adults' choice of AAC.
As a result, participants (4) will often avoid social situations rather than deal with strangers' reactions.

\subsubsection{Accommodating to current needs} 
\label{sec:current-needs}
Having a dynamic disability means that one's needs are not constant throughout their life or even throughout their day. Like any accessibility tool \cite{Shinohara_2011}, autistic adults need to find the right AAC for their current needs \cite{Donaldson2021}.
Here, there were themes of using energy to force speech or speak fluidly (4).
Switching to AAC can feel like more (1) or less (1) energy than using mouth words. One participant (1) would occasionally fall into old habits of speaking orally by default.
AAC can make it easier to overcome the ``\textit{initiation cost}'' of figuring out when to talk (1).

\citet{Zisk2019} note that intermittent speech--losing speech due to sensory stressors or co-occurring conditions--is rarely researched.
Our participants (9) mentioned losing the ability to use mouth-words because of various internal or external stressors.
Some (2) spoke about using AAC when they were overwhelmed or had a build up of stressors from throughout their day. Others (2) spoke about their co-morbid conditions that trigger the use of AAC.  Some participants (2) use AAC depending on their energy and the effort required.
It is important for users to have access to AAC tools and be able to identify when their bodies are reaching a tipping point so that they know when to start using AAC and select what AAC to use (2). One participant (1) will ask their partner to help them hold themselves accountable so that they respect their body's boundaries.

\subsubsection{Other conversational contexts} 
\label{sec:conversational-contexts}
\citet{Donaldson2021} have seen that autistic adults' AAC selection depends on the conversational context. Our interviews back this up as well:

\paragraph{Accommodating to the environment.}
Several participants (9) spoke about how the environment such as the weather or noise levels affects their AAC usage.
Participants (9) spoke about using their AAC tool to show text in situations where the environment prohibits TTS from being heard.
There are also situations in which participants do not have access to their device because of what they were doing (5), there was no room (2), or they were concerned that the device would be damaged -- e.g., around water (2).

\paragraph{Accommodating to the task.} 
Like other AAC users have found \cite{Obiorah_2021}, AAC is not always practical to use for every task.
A quarter of our participants (3) mentioned constraints with using AAC to give presentations to groups of people, where they would run into clunky interfaces or bugs. 
Other users (2) will switch to alternate ways to communicate that are appropriate for the task beyond specialized or high-tech AAC applications, such as using email for important work communications or laminated cards for making grocery lists.

\paragraph{Accommodating to conversation partners.} 
Our participants would also switch AAC types depending on who they were communicating with. This need to switch can stem from accommodating the conversation partner's disability (3), repairing if the partner cannot understand the TTS voice (3), or adapting to their partner's skill of using a particular modality or application (2).

\subsection{Benefits}
\label{sec:benefits}
Of course adults would not seek out AAC if there were not significant benefits. 
AAC gives users the ability to feel socially connected to others (3), with our participants making friends with other AAC users (2).
One participant even stated having AAC as a special interest\footnote{In the autism community, a special interest is a topic or hobby that deeply fascinates someone, and they try to learn all they can about it.}.
In this section we will talk about three main categories where people found AAC helped them beyond using it as a communication tool:
\textit{Organizing thoughts} (\S \ref{sec:organizing-thoughts}),
\textit{Identity} (\S \ref{sec:identity}), and
\textit{Self-accommodation} (\S \ref{sec:self-accommodation}).

\subsubsection{Organizing thoughts} 
\label{sec:organizing-thoughts}
\begin{displayquote}
``\textit{Sometimes I talk faster than I think and struggle to explain a complex idea, and AAC lets me collect my thoughts and write a clear message.}''
\end{displayquote}

\noindent Participants (8) mentioned various ways in which AAC helped them ``collect their thoughts''. Their AAC can be used to reference what they said in a previous conversation (1) or earlier in the current conversation (1), to overcome choice paralysis (1), or to draft messages to be shared later via text or online (2).

Some autistic people have difficulty determining when it is their turn in a conversation \cite{Ochi_2019}. One participant uses their AAC to type up what they would say, whether or not they will use it in a conversation. With having \textit{what} they want to say out of the way,  they can focus on \textit{when} to say it.

\subsubsection{Identity} 
\label{sec:identity}
Several of our participants (9) spoke about how AAC enabled them to communicate their needs and act independently. 
Using AAC has helped participants feel ``\textit{like a whole person}'' by letting them:
\begin{itemize}
    \setlength\itemsep{0em}
    \item Be understood more (2)
    \item Express their needs (3)
    \item Speak for themselves (3)
    \item Talk about emotional things or in emotional situations (3)
    \item Have the ability to communicate when it would otherwise be difficult (5) (``\textit{Just still being able to contribute even on a bad day.}'')
\end{itemize}
Subsequently, AAC use can become closely intertwined with a user's sense of self:
\begin{displayquote}
    ``\textit{Something which was really helpful to [trans people online was][...] being able to look at themselves using AAC as someone who is trying to deal with gender and how does that work when the world is neuronormative and assumes we speak as well?}''
\end{displayquote}

\subsubsection{Self-accommodation} 
\label{sec:self-accommodation}
Since communication is so core to our experience as humans, some of our participants have structured their entire lives around accommodating their AAC use.
One participant (1) has made it so that the majority of their ``\textit{heavy lifting conversations}'' are via email, and another participant (1) has structured their life around AAC such that they ``\textit{don't run into the edges very often anymore}''.
More than half of our participants (7) rarely even use mouth-words since starting AAC, except with people they are close to (2), around people who do not support its use (3), when they cannot access it (2), or when they need to communicate in a certain way, such as quickly or with emotion (1).
Participants also feel more confident speaking orally when they know that they have AAC as a backup (2).


\subsection{Access as an Adult}
\label{sec:access}

Due to many autistic adults finding and accessing AAC later in life, there is an additional overhead that autistic children do not run into---either because they are still a part of the US education system and therefore can more easily access speech-language pathologists (\textit{Institutional support}, \S \ref{sec:institutional}), or because their parents and caregivers take over the details regarding things like setup (\S \ref{sec:setup}) and pricing (\S \ref{sec:price}).

\subsubsection{Institutional support} 
\label{sec:institutional}
Opinions about support from speech-language pathologists (SLPs) or other institutions---such as government programs in place to help people who use AAC or who are disabled in general---were mixed.
Some participants (4) spoke about how speech therapists or SLPs were unhelpful or even abusive, hindering their ability to use AAC or find the right AAC for themselves.
\begin{displayquote}
``\textit{In my experience, [...] speech therapists were not helpful. [...] There's a lot they can do to acclimate folks to AAC better.
Understanding what it means to have part-time speech or unreliable speech, how it shows up, and how to work with it instead of against it---so that you're working on total communication.}''
\end{displayquote}

On the other hand, SLPs can create opportunities for AAC users. One participant had an SLP that let them try out an expensive AAC so that they could decide if it was right for them or not first. Another person (1) wished that there was more access to speech-language pathologists as an adult so that they might get more help for their AAC use, a barrier also seen by \citet{Zisk2019}. Furthermore, several participants (5) expressed concerns that could be addressed with an SLP, such as whether an AAC app exists with the features that they like, how to get started with a particular type of AAC, or figuring out if they like certain features.

Regardless, some participants (3) spoke about the need for AAC to be ``\textit{more accessible to more people}'', which can potentially be addressed via institutional supports.
A third participant (1) had found SLPs unhelpful, but they spoke a lot about benefiting from state-level support; they were able to obtain AAC technologies over the years without having to buy them themself.

\subsubsection{Setup effort}
\label{sec:setup}
\begin{displayquote}
``\textit{I would prefer an end-user program on the computer-- [...] There's an option to import boards, but where am I importing them from? [...] It's not transparent as to what format it's in. [...] And you can't import a Proloquo board to Speech Assistant. I wish there was a standard.}''
\end{displayquote}
Because many of our participants were introduced to AAC as an adult outside of institutional support, they had to find and setup their applications themselves. Once people get access to an AAC application, they spend a lot of time setting the app up and making boards to fit their needs (4). 
With so many complicated pieces, it is difficult for users to remember where everything is and what it can do (3).
It makes for a big overhead for starting AAC or moving to another AAC application (2).

\subsubsection{Pricing and inaccessible operating systems} 
\label{sec:price}
Some participants (4) spoke about being prevented from getting the AAC that they thought would best fit their needs because it was only available on an operating system that they did not have access to.
AssistiveWare's Proloquo suite of applications are only developed for iOS. Therefore, users would need to invest in an Apple product, which are typically more expensive than Android devices:
``\textit{I don't take my iPad with me [...] most of the time.
It was expensive, and I don't want to break it.}''

Similarly, most participants (10) spoke explicitly about how affordability played an important role in choosing an AAC application, a pain point echoed by \citet{Zisk2019}.
Participants will avoid paying full price for expensive applications if possible (5) 
and would gravitate toward applications that are free (4). 
Applications will occasionally charge additionally for text-to-speech voice options (3).

\subsection{Stumbling Blocks for Continued Use}
\label{sec:stumbling}

Once our participants found and setup their selected AAC applications, there were still major issues that they would encounter. These issues might discourage users from using particular features of the apps or might make them disregard the application entirely. Most of these concerns, with the exception of \textit{External support} (\S \ref{sec:external}), do not appear to be autistic-adult--specific and could be shared with other AAC-using populations. 
The other, more common stumbling blocks include \textit{Speed} (\S\ref{sec:speed}), \textit{Voice quality} (\S\ref{sec:voice-quality}), \textit{Limitations of showing text} (\S\ref{sec:text-limit}), \textit{Device restrictions} (\S\ref{sec:device-restrictions}), and \textit{Unreliability} (\S\ref{sec:unrelability}).

\subsubsection{External support} 
\label{sec:external}
Because our participants mostly lacked institutional support, customer support can be key to whether or not a user will continue using an application. Participants who experienced good customer support (2) felt that they can rely on the company to continually improve the app. On the other hand, participants who experienced poor customer support (3) will switch applications, if possible, or suffer in silence since their concerns are not being actively addressed by the developers.

\subsubsection{Speed} 
\label{sec:speed}
Speed has long been a concern in AAC use \cite{Newell_1998}. Like other adult AAC users \cite{McKillop2018,Dai2022}, 
nearly all participants (11) mentioned the issue of speed with AAC, wanting ``\textit{to be able to communicate faster}'' and keep up in conversations (3). Even patient conversation partners are not immune to distractions (2) and can sometimes forget what an AAC user might be responding to or get distracted by others in the conversation.
This issue of communication speed can stem from issues with locating button words on the board (6) or the input method that is being used (5).

\subsubsection{Voice quality} 
\label{sec:voice-quality}
Most participants (10) expressed their concerns with poor text-to-speech (TTS) quality, which is a common issue with AAC users \cite{McKillop2018}.
Seven participants (7) felt that the available TTS voices were often hard to understand or hard to hear in general.
Two participants (2) attempt to get around misunderstandings by changing the speed of the voice to be slower. Other misunderstandings can come from when their conversation partners are unable to process a synthesized voice (2)

A couple participants felt that the quality of the voices was simply too poor to use them (2) or were concerned about the ability to be expressive with the voice (2). 
Other TTS restrictions include not being able to use AAC to talk to people over the phone (3) or to play the voices loud enough (4).

\begin{figure}[t]
\includegraphics[alt={Figure 3 is a picture of a participant holding their UbiDuo device. The device has a slightly smaller than average sized keyboard with a small backlit screen above it. The screen can fold down, but it looks like it wouldn't help it be any thinner.},width=8cm]{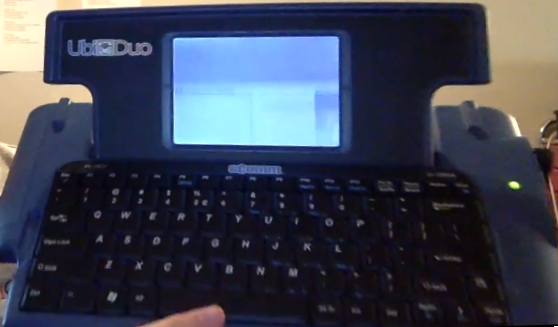}
\centering
\caption{A participant showing sComm's \textit{UbiDuo} device. These devices can be sold in pairs with either a wired or wireless connection. Despite its bulk, they prefers something like this over a phone app because of their ability to type quickly.}
\label{fig:bulky}
\end{figure}

\subsubsection{Limitations of showing text} 
\label{sec:text-limit}
Despite showing text being commonly used and even used as a back-up for TTS-based output, there are some severe limitations to it that auditory communication does not have.
Screens are often too small to show text (1) or can have a glare (1), which makes it particularly difficult to share with people at a distance (1) or with a group of people (2). Because of this, especially with phones, users have to hand their device to another person (2), which can be a security risk.

\subsubsection{Device restrictions} 
\label{sec:device-restrictions}
There are times when our participants (6) struggled with the physical devices themselves.
Smartphone screens are often too small to work for traditional AAC applications (3).
One of our participants (1) has a particularly interesting device called the UbiDuo---shown in Figure \ref{fig:bulky}, which has a standard keyboard and is not connected to the internet. However, it is incredibly bulky and single-purpose.
Even tablets are not perfect as older devices can have unpredictable quirks (2), such as crashing in extreme temperatures. 

\subsubsection{Unreliability} 
\label{sec:unrelability}
One point that was repeatedly brought up was the wish for continued, reliable use for AAC (6).
It is so important for one participant that they
``\textit{always have and always will carry pen and paper just because it's the most reliable}''.
It is especially crucial when even commonly-used applications have bugs. Several participants (6) have run into bugs while using their AAC. 
It can be disheartening, as well, when bugs are reported but the request to fix them is dismissed:
\begin{displayquote}
``\textit{I am by far not the only adult autistic AAC user who feels like Coughdrop support kinda blows us off.
Like, in the three months I used Coughdrop, I reported several bugs and *never* got any fixed, it was always "we'll forward it but no promises when he'll have time to look at it".} ''
\end{displayquote}

\subsection{Control of Communication}
\label{sec:control}

How companies log and use data has long been a concern within AAC research \cite{Pennington_2007,Reddington_2011}.
Wanting to be autonomous and in control of their own communication, autistic adults have concerns about people and companies potentially having access to everything they say. Many AAC applications are made with SLP intervention in mind, as children are learning to speak for the first time using AAC. Adults finding AAC later in life might not need or have access to these supports, and the conventional logging practices can seem invasive at best.

At the end of each interview, we listed features that loosely aligned with modern natural language processing techniques and asked participants whether they would use them, with the framing of privacy (never, only in the current conversation, across conversations, or always okay). The features were: automatic speech recognition, logging conversations verbatim, logging the topic of the conversation (i.e., topic modeling), saving personalized information automatically, saving personalized information explicitly, using a history of sentences said in the past (i.e., language modeling), and using a history of words said in the past (i.e., bag of words). The table that was shown to participants can be found in Appendix \ref{sec:appendix-privacy}, followed by the results of the responses in aggregate (Figure \ref{fig:privacy}).

\subsubsection{Control of features}
\label{sec:control-features}
While talking about privacy (starting at Question \ref{questions-privacy}), some participants (4) mentioned how they were okay using features as long as they are explicitly used and can be turned off easily. People do not want to be forced into certain features (1) and they want the ability to have control over details of how a feature is used (1).
\begin{displayquote}
``\textit{I don't like it when the way I use something automatically changes the way an app or website, or whatever, functions.
 I'm fine with it if I'm choosing to make the changes like deliberately adjusting settings or changing the layout of AAC app, for instance, but if it's automatically adjusting itself in response to how I interact with it, I hate that.}''
 \end{displayquote}
Further, half of our participants (6) said that automatic personalization was never a good feature, more than any other feature in our quantitative analysis (Figure \ref{fig:privacy}).

Some participants (4) would occasionally allow the logging of conversations as long as they had control over what, how, and when it was logging --- for example, if there was a way to delete items (1) or 
if the data was not analyzed in any way (1).
However, none of our participants were comfortable having logging verbatim always on (Figure \ref{fig:privacy}).
\begin{displayquote}
``\textit{I turned off every single prediction that ClaroCom [has], including its built-in support for learning automatically. I don't even like that it has those knobs. I'd rather it just plain didn't do any of that.}'' 
\end{displayquote}

\noindent Speech recognition also had 5 ``never'' ratings, with two participants (who both said ``never'') saying that they would be interested in that only if it was outside of an AAC application, and a third participant (``current'') would use it ``\textit{only if deliberately activated}''.

\subsubsection{Unwanted outsider access} 
\label{sec:unwanted-outsider}
Although some participants find company data collecting useful, participants do not want this collection to happen automatically. Most participants (11) were concerned about outsiders collecting their data without their consent. Mostly this concern was centered around companies logging data. While data security was certainly a concern, participants expressed explicit concerns about logs of conversations ending up in the wrong hands---such as people who hold control over the user (3). 

People (2) were also concerned about tracking data that they felt was unnecessary for an AAC app to track, such as location.
Other participants showed concern about people reading AAC text without their permission (3) 
or conversation partners looking through their phone (2):
\begin{displayquote}
``\textit{Sometimes my AAC use involves handing my phone to another person for them to read and I always have to trust that they're not going to then use my unlocked phone and go snooping.}''
\end{displayquote}

Only one participant (1) didn't express immediate concern about outsider access. They were the only participant who has been using various types of AAC since childhood, many of which were created pre-internet and only had data stored on the device. Now, their AAC use involves writing on paper and using services like Relay which have professionals reading their texts aloud:
\begin{displayquote}
``\textit{[Relay professionals are] held to really strict confidentiality standards, so I never have concerns about that.
[...] [they] can lose [their] license [...]
And if the sensitive information I would be sharing on Relay would be medical stuff, it's a stranger.
}''
\end{displayquote}

\subsubsection{Company trust}
\label{sec:company-trust}
Some participants (4) trust the company of their AAC applications well enough to handle their data.
Nonetheless, nine participants (9)
mentioned some uncertainty about the data policy of their AAC application, either in general or in the specifics of particular features of the app, and a couple participants (2) desire clear data policies.


\section{Research Directions}
\label{sec:specific-guidelines}

In this section, we highlight some of the technical aspects that our participants brought up which may  improve AAC but have not been, to our knowledge, mentioned much or at all in previous literature. 
Of course, these guidelines should be treated as such: guidelines. Any AAC applications that are being made should be prototyped and tested with their target user base. Furthermore, while these guidelines were discovered from autistic adults, we believe many of these would aid other populations of AAC users as well.

\subsection{Conversation Planning}
Some of our participants considered the slower speed a \textit{feature} of AAC, with participants using symbol-based AAC for calmer, more purposeful conversation (\S\ref{sec:symbol-text}). 
They use AAC to collect their thoughts and remember what they said previously in the conversation (\S\ref{sec:organizing-thoughts}), or to queue up things to say while they wait for a pause in the conversation (\S\ref{sec:current-needs}).
We recommend leaning into the unique capabilities of AAC by researching \textbf{ways to view conversation history, pull up previous conversation topics (summarized conversation history), or plan conversation topics}.
Furthermore, how can AAC aid with conversation planning without logging conversation history (\S\ref{sec:unwanted-outsider})?

\subsection{Keyboard Input}
Keyboard alternatives (e.g., \cite{Black_2019}), alternative keyboard mechanics (e.g., \citet{Wiegand2014}'s Swype-like interface), and other novel modalities for AAC have been extensively researched \cite{Ascari_2021,Curtis_2022,Curtis_2024}. There is, however, surprisingly little work on regular keyboard use with AAC applications \cite{Curtis_2022}.

All of our participants (\S\ref{sec:typing}) liked some form of keyboard functionality. Integrating a physical or on-screen keyboard made it so that users could leverage their typing speed or prevent the need to hunt for buttons. 
Proloquo includes an on-screen keyboard to allow the user to switch to typing in words, but only one word can be entered at a time. Some of our participants who used CoughDrop  integrated a keyboard into their home board (\S\ref{sec:typing}, Figure \ref{fig:embedded-keyboard}) but future work should look into \textbf{how AAC applications can intentionally integrate keyboards}. An initial step would be to include Bluetooth capabilities within AAC apps so that users can connect a physical keyboard, but future work should investigate \textbf{how to balance speed and convenience}.

\subsection{Opt-in Functionality}
Any feature that is not considered part of the core functionality of the application should be ``opt-in'' rather than ``opt-out'' (\S\ref{sec:control-features}). That is, users must know that the feature is there so that they can decide whether or not to turn it on. We need transparent data collection where users are told exactly what will be recorded \textit{only if} they opt-in to data collection in the first place. Potential research directions here include (1) \textbf{how to give the user control over when to collect data and/or remove certain sentences from what was collected}; or \textbf{how to best present and describe features so that users know they are available and know what they would do}. For example, visualization techniques might be useful for showing a user how their data is collected \cite{Hamidi_2018}, but how might this be extended to show \textit{what} was collected or what a feature does?

\subsection{Presentation Support}
A few participants mentioned the use case of giving presentations using their AAC (\S\ref{sec:conversational-contexts}). Due to the short, conversational form that AAC is usually designed for, users run into issues when using AAC for longer forms such as presentations. Research directions here might include:
\begin{enumerate}
    \item \textbf{Drafting paragraphs of text instead of single sentences}
    \item \textbf{Making tablets "project" TTS loudly enough or easily connect AAC applications to a speaker system}
    \item \textbf{Coordinating AAC applications with slideshow software}
    \item \textbf{TTS that is appropriate for emotive presentations (i.e., a presentation mode for TTS)}
\end{enumerate}

\subsection{Same-app Switching}
\label{sec:modular-uses}
Beyond customizing an AAC application for a specific user, it should enable a user with dynamic disabilities to switch between features, depending on their current needs, without changing applications.
We identify three examples that we believe would be worthwhile research directions for same-app switching. (1) \textbf{Easily switching between direct and indirect input} (\S\ref{sec:indirect}), (2) \textbf{simplifying the app's colors or board when the user is overwhelmed} (\S\ref{sec:customization}), and (3) 
combining symbol-based and text-based AAC within the same app (\S\ref{sec:symbol-text},\ref{sec:aging}). The application should integrate these in an intentional way that keeps the user in control (\S\ref{sec:control-features}) and does not impede motor plans (\S\ref{sec:typing}).

Separate from \textit{Keyboard Input}, \# looks at the ability to switch between symbol- and text-based AAC depending on the user's needs.
We have not found any evidence of research or products that attempt to \textbf{switch between the two separate modalities}. Integrating them into a single app could save the user time since they would not need to switch applications, and it would give them access to the same saved phrases across modalities.

\subsection{Trans* Voices}

While wanting personalized voices is not unique to autistic adults, there is a large overlap between trans* and autistic populations \cite{Warrier_2020}, making gender-diverse speech synthesis valuable to the autistic community. Our participants want
inclusive and diverse text-to-speech (TTS) voices so that they can find a voice that matches their \textit{identity}, not necessarily their body's voice (\S\ref{sec:voice-identity}).
Research should continue to look at creating voices not based on voice cloning and understanding how trans-feminine, trans-masculine, nonbinary, and genderqueer people sound and/or want to sound with TTS.
There is some work on creating gender-expansive voices for TTS \cite{Hope_2020,Székely_2023,Hope_2024,Szekely_2024}.
In addition to improving these voices themselves, future research needs to look at \textbf{how people feel when using these voices for their daily AAC} and \textbf{how to create TTS voices that can be customized by trans users}.


\section{Limitations \& Future Work}
Although we were able to collect some variation in gender and age, all of our participants were white. This is a significant limitation to our work, which we attribute to our recruitment methods. Therefore, while our insights might apply to BIPOC users, they would have to be validated in future studies. This is particularly important since some \cite{Pope2022} have identified racial disparities in access to such communication tools.

Another limitation is that our participants were only recruited online. As we did the majority of our interviews during the COVID-19 global pandemic, there was not much of a choice. Because of this we are missing out on a fraction of the autistic adult population that does not use the internet or social media. This could be why our participants skewed younger.

We also recognize that only 12 people were interviewed in this study. While we were able to get an in-depth look into some participants' use of AAC, this only provides a snapshot of the types of design decisions we might want to consider when creating AAC applications. A much larger, shallower study must be done in order to understand a variety of perspectives.

Many existing AAC applications are created for children first, leaving autistic adults to force applications to fit their needs. Although we are certainly not the first to suggest an adult-centered design \citet{McNaughton_2007}, 
we hope that this work is just the beginning of an emphasis on adult-first design for autistic adults and people with other developmental disabilities.
Specifically, we would like to see AAC design that 1) treats users as adults or helps users be treated as adults, and 2) supports users who do not have access to an AAC specialist.
Adult-centered AAC might include features such as no recording of button presses (\S\ref{sec:aging},\ref{sec:unwanted-outsider}), integrating an "adult" vocabulary (\S\ref{sec:vocab},\ref{sec:aging}), developing more expressive adult TTS voices (\S\ref{sec:voice-identity}), or creating symbol libraries that feel more mature (\ref{sec:aging}).
Supporting users without SLP access does not have to be constrained to social scaffolds and could be integrated into the design of the applications themselves. For example, they could include features for making setup easier (\S\ref{sec:setup}) or helping people figure out what type of AAC tool would best fit their needs (\S\ref{sec:institutional}).


\section{Conclusion}
Our interviews with autistic adult AAC users revealed both current societal and technical issues encountered with AAC applications. 
We identified what types of input \& output options people prefer for their AAC, what considerations people take into account for when selecting an AAC application or switching to a different AAC application, the benefits of using AAC, issues with accessing AAC as an adult, and concerns about control over their communication.
We concluded by suggesting seven novel research directions for AAC researchers to make AAC better suited specifically for autistic adults.
\begin{enumerate}
    \item \textit{Conversation Planning} - Referring to what the user has previously said in the conversation
    \item \textit{Keyboard Input} - Typing integration
    \item \textit{Opt-in Functionality} - Practices for making features opt-in but still informing users that they're there and what they do
    \item \textit{Presentation Support} - Features that enable users to give long-form presentations to larger audiences
    \item \textit{Same-app Switching} - Ways of switching between input types, interfaces, or modalities within the same app
    \item \textit{Trans* Voices} - Customizable and gender-expansive synthesized voices
\end{enumerate}
We would like to encourage the rest of the AAC research community to consider our insights and suggested research directions so that we can iterate on them as a community.

\begin{acks}
This work was funded in part by the National Science Foundation under Grant \#2030859 to the Computing Research Association for the CIFellows Project. We would like to extend a special thanks to Dr. Layne Jackson Hubbard and Dr. Alyssa Zisk for looking over the interview questions; Dr. Dana{\"e} Metaxa \& Dr. Andrew Head for HCI methods help; and Dr. Chris Callison-Burch for being a great postdoc/masters mentor. Extra special thanks to all of our participants! We hope we did your words justice.
\end{acks}

\bibliographystyle{ACM-Reference-Format}
\bibliography{refs}

\appendix

\section{Pre-Interview Screening Questionnaire}
\label{sec:appendix-intake}
\texttt{The following is the text shown to potential participants via a Qualtrics survey. Any text in this section that uses this font was not displayed to the user and is only included in this paper for additional context.}

Hello! Researchers at the University of Pennsylvania are currently looking for autistic people to be interviewed about their AAC (Augmentative and Alternative Communication) usage. 
Please note that this study is on digital AAC applications, such as those used on phones, tablets, or other devices.
If you are interested in being interviewed for our study, please fill out this preliminary survey to see if you qualify. Thank you!

\begin{enumerate}
    \item What is your email address? If you qualify for the interview, we will contact you via email. \texttt{[text entry]}
    \item Would you like to participate in our study?
        \begin{enumerate}
            \item Yes
            \item No
        \end{enumerate}
    \item Are you autistic? (self-diagnosed is fine)
        \begin{enumerate}
            \item Yes
            \item No
            \item Maybe
        \end{enumerate}
    \item What is your age?
        \begin{enumerate}
            \item Less than 18 years old
            \item 18-24 years
            \item 25-34 years
            \item 35-44 years
            \item 45-54 years
            \item 55-64 years
            \item 65-74 years
            \item 75 years or older
        \end{enumerate}
    \item Which country do you \textbf{currently} live in?
    \texttt{[Drop-down list of countries]}
    \item Are you a native English speaker? (Have you been speaking English since childhood?)
        \begin{enumerate}
            \item Yes
            \item No
        \end{enumerate}    
    \item Are you non-speaking (i.e., do not speak verbally) or lose speech occasionally?
        \begin{enumerate}
            \item Non-speaking
            \item Mostly non-speaking
            \item Selectively mute/situationally mute
            \item Semi-speaking
            \item Unreliable speech
            \item Other AAC user \texttt{[with free-form text entry]}
            \item None of the above
        \end{enumerate}    
    \item Do you use an AAC application to communicate part-time or full-time?
        \begin{enumerate}
            \item No
            \item Part-time
            \item Full-time
        \end{enumerate}    
\end{enumerate}

\texttt{If the participant selected ``Part-time'' or ``Full-time'', they would be shown the following questions:}
\begin{enumerate}[resume]
\item Approximately how long have you been using AAC for?
        \begin{enumerate}
            \item Since childhood
            \item 10+ years
            \item 5-9 years
            \item 2-4 years
            \item 1 year or less
        \end{enumerate}
\item How long have you been using your current AAC app?
        \begin{enumerate}
            \item Since childhood
            \item 10+ years
            \item 5-9 years
            \item 2-4 years
            \item 1 year or less
        \end{enumerate}
\item What do you look for in an AAC application in general? (There are no right or wrong answers!) \texttt{[text entry]}
\item Do you have any questions or comments for the researchers? \texttt{[text entry]}
\end{enumerate}

\texttt{As we were focusing on autistic adult AAC users who spoke English, any participants who reported ``No'' for question 2, ``No'' for question 3, ``Less than 18 years old'' for question 4, ``No'' for question 6, or ``No'' for question 8 were given an automated rejection notice. The rest of the participants were shown the message:}

Thank you for your interest in our study!
We will reach out to you shortly to schedule an interview. Please expect an email from \texttt{[redacted email address]}.

\section{Interview Questions}
\setcounter{footnote}{0} 
Below are the questions that the participants were asked. Section \ref{sec:appendix-speaking} shows the entire set of interview questions verbatim, including text formatting such as bolded text, underlining, and a footnote. 

Mostly speaking and mostly nonspeaking participants were provided the same questions with the exception of Question 1. Section \ref{sec:appendix-nonspeaking} provides the alternative Question 1 that we asked participants who were primarily nonspeaking.

Once the participants had looked at all of the questions, they were allowed to fill out the privacy table (Section \ref{sec:appendix-privacy}) that was given in a separate PDF called `wait\_to\_open.pdf'.

\subsection{Interview Questions (with Question 1 for mostly speaking participants)}
\label{sec:appendix-speaking}
\textbf{For any question, if something doesn't apply to you, please explain.\\
All of the questions relate to dedicated AAC applications and related tools you might use (such as word processors).}

\begin{enumerate}
    \item Think of a recent time when you used AAC.

    \begin{enumerate}
        \item What triggered your need to use it (if you’re comfortable with sharing)?
        \item How did the conversation go? Were there misunderstandings?
        \item Approximately how long did you use it in the single conversation?
        \item If you speak orally as well, what are ways AAC has supplemented your conversations?
        i.	Are you using AAC alongside or instead of speech?
        
        \item What other ways did you communicate (e.g., non-verbal communication)?
    \end{enumerate}

    \item The following questions will ask you to think of specific situations.
    \begin{enumerate}
        \item Tell me about a situation where you wanted to use AAC but \textbf{couldn't} because of the \underline{limitations of the application(s)}.
        \item Tell me about a situation where you wanted to use AAC but \textbf{couldn't} because of the type of \underline{social situation}.
        \item Do you use AAC in private, out in public, or both?

i.	If you haven’t already, tell me about a time you used it in private.

ii.	If you haven’t already, tell me about a time you used it in public.

iii.	How did these two times differ? Was one easier than the other?

\item Tell me about a time when you used AAC but \underline{it didn't work}. 

i.	For example, the AAC application was not sufficient for your communication needs or even the application itself was broken in some way.

ii.	Describe some things that you dislike(d) about the app/tool.

iii.	Have you continued to use the app/tool? Why or why not?

    \end{enumerate}

    \item Describe a typical day when you use AAC. 
    
    [This is the part where I’d like pictures showing me your application(s). It can be a screencap or a picture taken from another device.]

\begin{enumerate}

\item What app(s) do you use?

\item If you have multiple AAC apps, how do you decide which one to use?

\item Can you show me or tell me about what you like about your app(s)?

\item Is there anything that you wish the app could do that it doesn’t do?

\item Do you think this AAC app limits you in some way? If so, how?

\end{enumerate}

    \item Empowerment \footnote{“the process of becoming stronger and more confident, especially in controlling one's life and claiming one's rights” – Oxford Dictionary}
    \begin{enumerate}
\item Tell me about a time AAC helped you feel empowered as an autistic person.

\item Tell me about a time AAC made you feel \textbf{disempowered} as an autistic person.

i.	Now imagine a brand new AAC tool that could have helped you in that moment. What would it be like?

\item How do you imagine future AAC \textbf{could} empower you as an autistic person?

\end{enumerate}

    \item Think about the types of AAC applications you have used (both now and in the past).
    \begin{enumerate}
\item What types of features do you look for in an AAC app or tool?

\item What features do you try to avoid?

\end{enumerate}

    \item Have you ever been concerned about your privacy when using AAC?

\begin{enumerate}
\label{questions-privacy}
\item What type of information about you do you think your AAC application(s) saves?

\item Where do you think it’s stored (e.g., on your device, in “the cloud”)? 

i.	How secure do you think it is?

ii.	How long do you think it’s stored for?

\item Is there any feature that you would never allow to be in an AAC application you use? Any that you do allow but would prefer not to have?

\item Once you answer Question 6c, open the file ``wait\_to\_open.pdf'' and rate how you feel about the features in the table.

\end{enumerate}

    \item Is there anything else you think I should know that we haven’t spoken about?
\end{enumerate}

\subsection{Alternate Question 1 for Mostly Nonspeaking Participants}
\label{sec:appendix-nonspeaking}

\begin{enumerate}
    \item Usage.
    \begin{enumerate}
        \item When do you \textbf{not} use AAC? 
        \item Think of a recent conversation you've had. Were there misunderstandings because of AAC?
        \item If you speak orally as well, what are ways AAC has supplemented your conversations?
        
        i.	Are you using AAC alongside or instead of speech?
        \item What other ways did you communicate (e.g., non-verbal communication)?
    \end{enumerate}
\end{enumerate}

\begin{table*}[h!]
\begin{tabular}{ |p{.08in}|
   p{2.2in}|
   p{.4in}|
   p{.8in}|
   p{.8in}|
   p{.4in}|}
\hline
 \textbf{\#} & \textbf{AAC Feature} & \textbf{Never} \newline \textbf{Allow} &\textbf{Current \newline Conversation} & \textbf{Across \newline Conversations} & \textbf{Okay} \\\hline
 1 & using speech-to-text
(i.e., having the app “listen” to people around you)	 &  &  &  &\\	\hline		
2 & keeping a log of what has been said in the conversation word for word 				 &  &  &  &\\\hline
3 & keeping a log of the topic(s) of the conversation		 &  &  &  &\\	\hline	
4 & saving personalized information (e.g. favorite TV show, best friend’s name) \textbf{automatically}				 &  &  &  &\\\hline
5 & saving personalized information \textbf{only when you explicitly permit it	}			 &  &  &  &\\\hline
6 & using \textbf{sentences} you've said in the past to predict what you want to say				 &  &  &  &\\\hline
7 & using \textbf{words} you've said in the past to predict what you want to say		 &  &  &  & \\\hline
\end{tabular}
\end{table*}

\newpage
\subsection{Privacy Table}
\label{sec:appendix-privacy}

For each row of the table, read the feature and think about how you would feel using AAC with that feature—think specifically in terms of \underline{your privacy}.
The options are:
\begin{itemize}
\item \textbf{Never Allow} - This is a feature you would never permit in AAC. It might even discourage you from choosing/using the app.
\item \textbf{Current Conversation} - This is a feature that you’re okay with as long as the data is used in the current conversation only and then deleted.
\item \textbf{Across Conversations} - This is a feature that you’re okay with across several conversations with the assumption that the data will be deleted at some point in the future—for example, either by you or by the app after a certain period of time.
\item \textbf{Okay} - This is a feature that you’re always okay with, regardless of the circumstances.
\end{itemize}

Put an X in the cell that best matches your preference for each feature. Feel free to elaborate on why or how you made your choices.

\begin{figure*}[bh]
\includegraphics[alt={Number of people who would use each feature. People were the most comfortable with deliberate personalization, followed by word prediction and sentence prediction. Automatic personalization was the least liked, followed by speech recognition. Logging (verbatim or topics) were both in the middle. Noteably though, nobody said they were always okay with logging verbatim.},width=\textwidth]%
{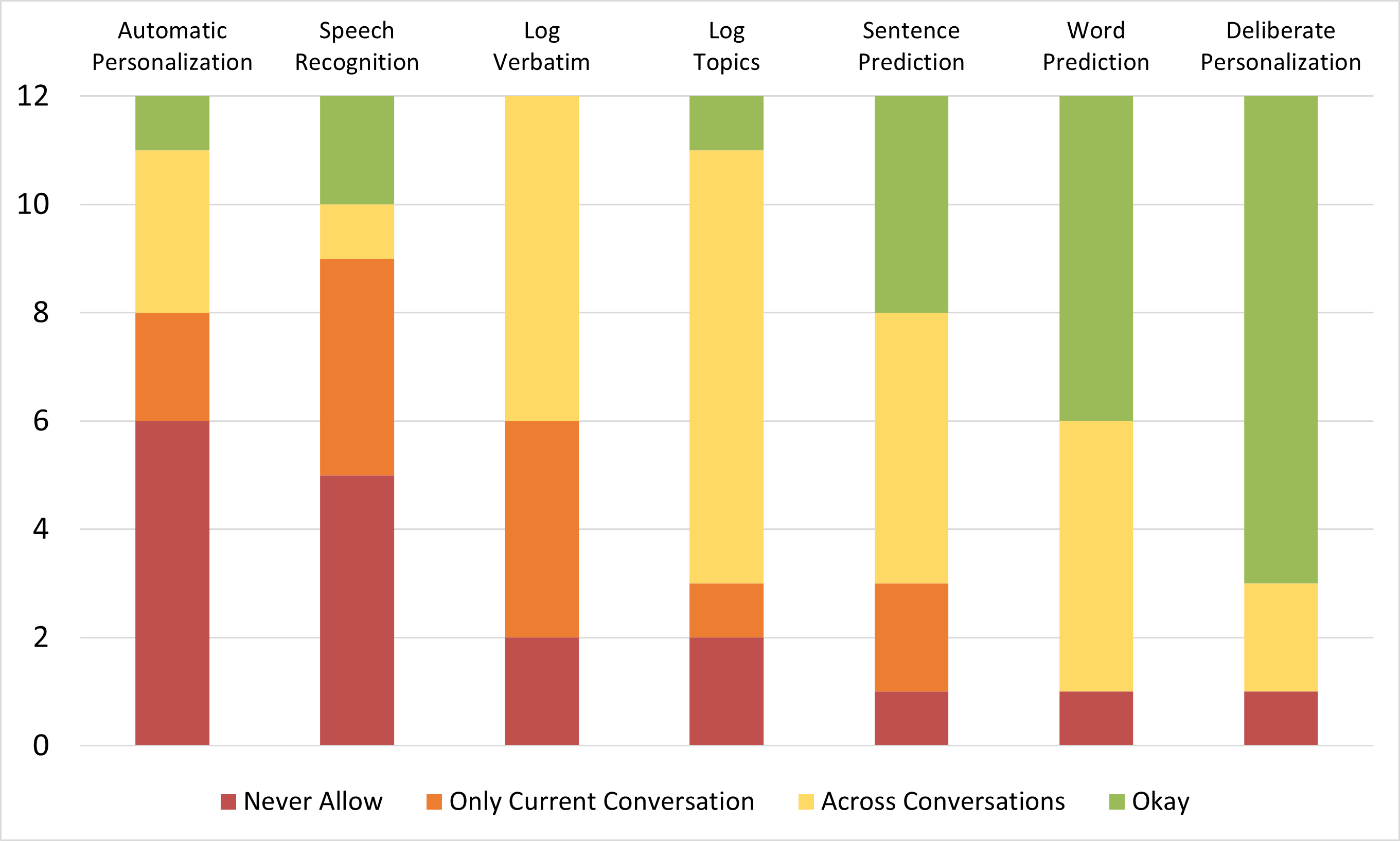}
\centering
\caption{Privacy Table Results -- Level of comfort for use of and data storage from 7 potential features. Participants rated each feature as either something they would never use (red), something they would only use as long as the data persisted only in their current conversation (orange), something they would use with the data persisting across conversations --- with the assumption that it would be deleted over time (yellow), or something they would be okay using all the time (green). The specific descriptions for each feature can be found in Appendix \ref{sec:appendix-privacy}.}
\label{fig:privacy}
\end{figure*}

\end{document}